\documentclass[12pt]{article}
\usepackage{pdproc,epsfig} 

\begin{document}

\renewcommand{\thefootnote}{\alph{footnote}}
  
\title{
MEASURING EARTH MATTER DENSITY AND TESTING THE MSW THEORY~\protect\footnote{
Written version of a talk presented at the 
``XII-th International Workshop on Neutrino Telescope'', Venice, Italy, 6-9, March 2007.}} 

\author{HISAKAZU MINAKATA}

\address{Department of Physics, Tokyo Metropolitan University, \\
1-1 Minami-Osawa, Hachioji, Tokyo 192-0397, Japan\\
 {\rm E-mail: minakata@phys.metro-u.ac.jp}}

%  \centerline{\footnotesize and}

%\author{SECOND AUTHOR'S NAME}

%\address{Group, Company, Address, City, State ZIP/Zone,Country}

\abstract{
In this talk I have raised the question of how the future discovery of leptonic 
CP violation can be made robust even at accepting the rather large current 
experimental uncertainties in our knowledges of neutrino propagation in matter. 
To make progress toward answering the difficult question, 
I listed ways to proceed: 
(1) Obtain tighter constraints on the MSW theory by testing it by 
various neutrino experiments. 
(2) Measure the matter effect {\em in situ}, namely within the experiment for 
discovering CP violation itself.  
(3) Uncover leptonic CP violation in a matter effect free environment. 
I also reported a step made toward the above point (2) by taking neutrino 
factory as a concrete setting; 
An accurate {\em in situ} measurement of the matter effect looks promising. 
}
   
\normalsize\baselineskip=15pt

\section{Measuring matter density for discovering CP violation?}

Leptonic CP violation may be one of the most important elements 
of our fundamental understanding of matter. 
Most probably it is related to baryon number asymmetry in our 
universe \cite {leptogenesis,talks}.  
Therefore, its exploration is one of the most serious targets of 
future neutrino experiments. 
I argued in the last year in Venice \cite{NOVE06_mina} that leptonic 
CP violation should exist both in the form of 
Kobayashi-Maskawa type \cite{KM} and 
Majorana type \cite{Mphase} phases. 
It is my pleasure to be here again, by the kind invitation by Milla, 
to further pursue the line of thought. 
I know that you may say that the title of this talk does not sound 
right for the topics of CP violation. 
But, in fact, the motivation of my talk addressing this topics {\em does} 
come from the question of how CP violation (of Kobayashi-Maskawa type) 
can be uncovered experimentally.

Let me start by describing my motivation, a personal one. 
In a recent work \cite{T2KK1st,T2KK2nd} we have described 
a concrete proposal for measuring the unknown parameters 
in the lepton mixing and determining the neutrino mass pattern  
by placing two identical megaton-class detectors 
one in Kamioka and the other in Korea which receive an intense 
neutrino beam from J-PARC facility. 
It may be called as the Tokai-to-Kamioka-Korea setting, which is 
sometimes dubbed as ``T2KK'' \cite{T2KKweb}. 
We have demonstrated that by using the setting one can resolve 
the eight-fold parameter degeneracy, not only the one related to 
the mass hierarchy and CP phase \cite{intrinsic,MNjhep01}  
but also the $\theta_{23}$ degeneracy \cite{octant},  
if $\theta_{13}$ is within reach by the next generation accelerator 
\cite{T2K,NOVA} and reactor experiments \cite{MSYIS,reactor_white}.

But, one day, I asked myself; 
``Is this way of uncovering leptonic CP violation robust?''. 
Robustness that I concern implies robustness against possible change 
in the basic framework that may arise due to the lack of our knowledge. 
Of course, full treatment of the whole arbitrariness allowed at this 
moment is just too far a goal, if not impossible, to reach. 
For example, if the three neutrino mixing is too tight to accommodate 
what is happening in nature we stuck, or more appropriately, 
we face with the situation that a gigantic number of completely different 
scenarios of how CP is violated are possible.

Therefore, I have to be more specific about the setting of the problem. 
What I mean robustness in this talk refers to the one to uncertainties 
in our understanding of theory describing neutrino propagation in matter, 
the MSW theory \cite{Wolfenstein,MS}. 
More specifically, I concern uncertainties in the MSW coefficient $c_{MSW}$, 
which is defined by replacing the standard matter potential 
$a \equiv \sqrt{2} G_{F} N_{e}$ by $c_{MSW} a$. 
They may arise through any effects which renormalize the MSW coefficient. 
An unimaginative list for such effect includes: 
(1) uncertainty in matter density measurement in the earth,  
(2) presence of non-standard neutrino interactions \cite{Wolfenstein}. 
I warn you that the list might be short simply because of our ignorance. 
It is quite possible that there is a real candidate for (2) if a new physics 
at TeV scale is waiting for discovery.

If the effects outside the standard electroweak theory are absent, 
looking for the deviation of $c_{MSW}$ from unity can serve as 
a pure test of the MSW theory.\footnote{
%%%%%%%%%%%%%%% \footnote %%%%%%%%%%%%%%%%%
One may argue (as some people in the audience did) that the 
MSW theory cannot be in error because it is based on standard 
electroweak theory. 
While it is very likely to be the case, I want to emphasize that it is 
important to confirm our understanding of the theory in the region of 
coherent forward scattering. 
Recall that despite people believe in Einstein's theory of general relativity, 
yet, they still continue to perform various experimental tests to confirm 
the correctness of the theory. 
Let me remind you that we never saw a spoon is bent by neutrinos. 
Or, more precisely speaking we saw the spoon may be bent, but only 
at angle resolution with $\sim$100\% error. 
}
%%%%%%%%%%%%%%% \footnote %%%%%%%%%%%%%%%%%
%
It would be desirable if we could distinguish these two aspects, 
in and outside the electroweak theory,  in testing the MSW theory. 
But, in general their effects mix with each other and both act as 
ambiguities in estimating the background to CP violation search. 
If non-standard neutrino interactions have richer flavor structure 
it might be possible to separate between these two effects.

I believe that it is the very relevant issue not only in T2KK but also 
in many other approaches which seek to uncover CP violation. 
It is because leptonic CP violation is severely contaminated 
by the matter effect. 
(For early references on the CP phase-matter effect interplay, 
see e.g., \cite{AKS,MNprd98}.) 
If the MSW theory is in error, or neutrinos have 
non-standard flavor changing neutral current interactions, 
many discussions on how to separate genuine CP violation 
due to the CP violating phase from the fake one by 
the matter effect, most probably, ruin.
The potential fragile feature of the method for discovering leptonic 
CP violation is related to the absence of model-independent 
(or framework independent) measure for CP violation, lepton analogue of 
``$K_{L} \rightarrow 2 \pi$'', 
an unmistakable clean signature which cannot be masked by 
competing fake effects.\footnote{
%%%%%%%%%%%%%%% \footnote %%%%%%%%%%%%%%%%%
One can argue that T violation measurement can provide such 
clean signature because it is not obscured by the matter effect 
\cite{krastev-petcov}. 
While it is in principle true, the experimental setup which is required 
to embody the clean feature of T violation measurement is very demanding. 
At this moment no concrete proposal for such setup is available.
}
%%%%%%%%%%%%%%% \footnote %%%%%%%%%%%%%%%%%

You may say that 
``Oh, the MSW theory is already verified by the solar neutrino experiments''. 
Then, I ask ``In what sense and to what accuracy?''. 
Let me first discuss this question before entering into my problem.

\section{Is MSW theory verified by solar neutrino observation?}

After decades of struggle in solar neutrino observation pioneered by 
Davis \cite{Davis}, the solar neutrino problem is now solved. 
Among other solar neutrino experiments \cite{solar}, 
SNO finally confirmed that solar neutrinos experience  
flavor transformation by its {\em in situ} measurement of CC/NC ratio \cite{SNO}, 
the phenomenon first discussed by 
Maki, Nakagawa and Sakata \cite{MNS}. 
Then, KamLAND pinned down the nature of the phenomena as 
due to mass-induced neutrino oscillation \cite{KamLAND}. 
It verified that the phenomenon, which was first discovered in the 
atmospheric neutrino observation by Super-Kamiokande \cite{SKatm}, exists 
also with the solar $\Delta m^2$ scale.

The resultant solution, so called the large mixing angle MSW solution, 
has a unique feature, which is important in our context. 
The mixing parameters selected out live deep inside the adiabatic region. 
Then, the electron neutrino survival probability can be written under 
the approximation of small $\theta_{13}$ as 
\begin{eqnarray}
P_{ee} = \frac{1}{2} + 
\frac{1}{2} \cos 2\theta \cos 2\theta_{m},
\label{Pee}
\end{eqnarray}
where $\theta$ and $\theta_{m}$ stand for the mixing angle in vacuum 
and in the solar matter, respectively. 
If we further assume that neutrinos are produced in much deeper 
region than resonance density in the sun, we can make further approximation 
$\theta_{m} \simeq \pi/2$. Then, 
$P_{ee} = \sin^2 \theta$. 
That is, the $\nu_{e}$ survival probability can be expressed only by 
the vacuum parameter.\footnote{
%%%%%%%%%%%%%%% footnote %%%%%%%%%%%%%%%%%
The fact that $P_{ee}$ can be written only by the vacuum parameter 
should not be misunderstood as absence of the matter effect. 
Rather, it is the ``matter effect dominated'' situation in which $P_{ee} $ 
can become much less than 1/2.
}
%%%%%%%%%%%%%%% footnote %%%%%%%%%%%%%%%%%
%
Of course, there is a matter density dependent correction. 
While it is not so small, but it does not appear to be detected by the 
current experiments.

Therefore, $P_{ee}$ at high-energy region of $^8$B neutrinos depends 
only weakly to the matter density in the sun. 
It is a good news and at the same time a bad news.
It is a good news because the theoretical prediction of the $^8$B neutrino 
rate depends in a sensitive manner neither on the absolute matter 
density in the sun, nor on details of the matter density profile in the sun, 
the celebrated robustness of the prediction by the LMA MSW solution. 
On the other hand, it means that observation of $^8$B neutrinos may not 
be used as a sensitive tool for an accurate test of the MSW mechanism. 
This is the basic reason why the solar neutrino observation can 
constrain the MSW coefficient only up to a factor of 2 uncertainty 
\cite{bari_focus}. 
(Recall that the MSW coefficient $c_{MSW}$ is defined by replacing the 
standard matter potential $a \equiv \sqrt{2} G_{F} N_{e}$ by $c_{MSW} a$.)

While it is an independent question, 
I want to recall that the LMA MSW solar neutrino solution have not yet 
been confirmed in a manner independent of the standard solar model. 
Its characteristic predictions, the spectral upturn of the $^8$B 
neutrinos at low energies and the day-night variation of neutrino flux of 
about 2\%, despite people's great effort, have never been seen.\footnote{
%%%%%%%%%%%%%%% footnote %%%%%%%%%%%%%%%%%
If one trusts relative normalization between $^8$B and low energy $pp$ plus 
$^7$Be neutrino fluxes of order $\sim10^4$ predicted by the standard 
solar model, then one can show that there is an evidence for spectrum 
upturn by comparing between the SK-SNO and the Ga data. 
}
%%%%%%%%%%%%%%% footnote %%%%%%%%%%%%%%%%%

To summarize, I have to conclude at this stage that, 
despite the strong evidence for the presence of matter effect 
in solar neutrino observation, the MSW theory in its current status 
is not established experimentally at the sufficient accuracy to be 
used for reliable estimation of the background matter effect in future 
CP violation search.

\section{What to be done?}

What I want is simply to obtain robust evidence for leptonic CP violation.
By ``robust'' I mean 
``remain valid even after fully taking account of any experimental 
uncertainties that still exist in the framework I use to define the 
measurement of CP violation''. 
I know that it is not an easy goal to make.  
But, I believe that discovery of CP violation can become a truly  
experimental statement only when it is done.

Let me try to describe some considerations to make progress 
toward the goal. 
Let me restrict myself into the experimental uncertainty in verifying 
the MSW theory. One can proceed along one of the following ways:

\begin{itemize}

\item

Prove or obtain much tighter constraints on the MSW theory by 
testing it by various neutrino experiments.

\item

Measure the matter effect felt by neutrinos {\em in situ}, namely within 
the experiment for discovering CP violation.  

\item

Uncover leptonic CP violation in a matter effect free environment.

\end{itemize}

The last possibility sidesteps the problem of uncertainty in experimental 
verification of the MSW theory in uncovering CP violation; 
It is certainly a good way to proceed. 
The project which is best suited for this purpose would be the 
CERN-MEMPHYS project \cite{MEMPHYS}. 
(The next closest one may be T2K II \cite{T2K}.) 
Personally, however, it is not my choice because of possible drawback 
of this approach; 
It is unlikely that the same experiment can determine the mass hierarchy 
(unless one utilizes the alternative channels, such as high-statistics 
atmospheric neutrinos or supernova neutrinos, etc.). 
Another relatively matter effect free way of detecting CP violation 
would be the accelerator-reactor combined method \cite{reactorCP}.

Unfortunately, I am not able to solve the problem that I raised. 
The purpose of my talk today is merely to pose it in a correct way  
with expectation that people can make progress toward solving the issue.

\section{Neutrino factory measurement of CP violation}

The question I raised might sound too hard to solve. 
But, the situation is not that bad. 
Let me try to illustrate this point by taking a concrete example.

The problem of matter effect contamination in uncovering CP violation 
is most serious in neutrino factory \cite{nufact}. 
In a standard setting of placing detector at 3000-4000 km from a 
muon storage ring \cite{golden}, as is well known, 
the matter effect dominates over the CP phase effect. 
See, e.g., Fig.~8 in \cite{MNjhep01} for illustration of this fact 
by the bi-probability plot. 
Despite the undesirable feature, people coined into the neutrino factory 
because muon detection is extremely clean, 
in particular at high energies.\footnote{
%%%%%%%%%%%%%%% footnote %%%%%%%%%%%%%%%%
An alternative strategy discussed at the similar stage of understanding 
how to measure CP violation was low-energy superbeam \cite{superbeam}. 
These two approaches are contrasted  in \cite{NOW00_mina}. 
}
If $\theta_{13}$ is extremely small, $\sin^2 2\theta_{13} \ll 0.01$, 
it is the leading candidate for the machine entitled to search for 
the ``diamond'' in the frontier. 
An alternative approach is based on ``beta beam'' \cite{beta}.

So far, the problem of ambiguity to CP phase measurement due to the 
matter effect has been addressed in a limited sense, in a form of 
uncertainty in measured matter density by the geophysical method. 
See e.g., \cite{intrinsic,Joe,yasuda,munich1,Ohlsson,munich2}. 
Here, the problem I address is much more broad; 
In addition to the error of geophysical matter density measurement, 
it includes uncertainty in the theoretical framework in a range that 
is still unconstrained by the current (or the future) experimental measurement.

\section{ {\em In situ} measurement of the MSW coefficient by neutrino factory}

The fact that the problem is most severe in neutrino factory 
is, in a sense, a ``good news'' because then one has to necessarily 
solve the problem. 
Otherwise, one cannot make the goal of unambiguous demonstration 
of CP violation. 
In fact, we recently made a concrete proposal to solve the problem 
{\em in situ} in measurement in neutrino factory \cite{mina-uchi}. 
Let me describe our proposal in this section.

\subsection{Which baseline? }

\begin{figure}[h]
\vspace*{0.5cm}
\begin{center}
\epsfig{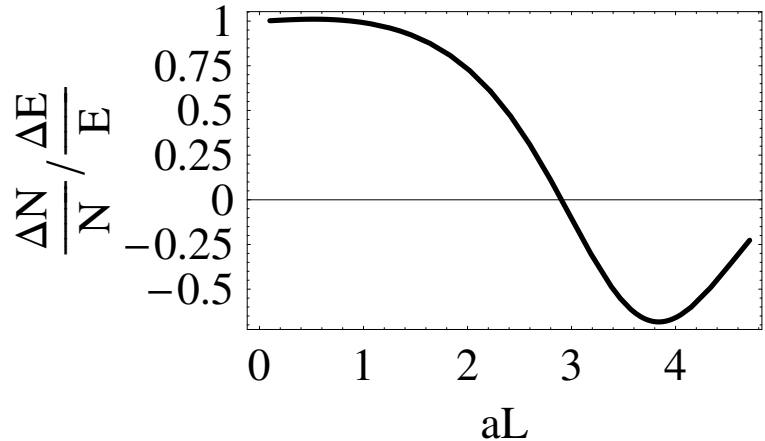}
\end{center}
\vspace*{0.2cm}
\caption{ $aL$ dependence of the double ratio 
$\left( \Delta N / N \right) / \left( \Delta E / E \right) $ defined in the text. 
Taken from \protect\cite{uchinami}.  }
\label{double-ratio}
\end{figure}

We have started from a general question at which baseline distance 
the matter density can be measured most accurately.\footnote{
%%%%%%%%%%%%%%% footnote %%%%%%%%%%%%%%%%%
When we talk about the measurement of matter density by neutrinos 
what we mean is, of course, the electron number density in the earth. 
Since the electron fraction $Y_{e}$ is very close to 1/2  in the earth 
it can be related to the matter density. 
}
To answer the question it is natural to consider the response to change 
in energy because the ratio of the matter effect to the vacuum effect, 
$aL / \Delta_{31} \equiv  
\left( \frac{1}{\sqrt{2}} G_F N_e L  \right) /  
\left( \frac{|\Delta m^2_{31}| L}{4 E} \right) $, 
is proportional to neutrino energy $E$. 
If we measure number of events $N(E)$ and $N(E + \Delta E)$ 
at two different energies $E$ and $E + \Delta E$, 
we obtain the double ratio 
$\left( \Delta N / N \right) / \left( \Delta E / E \right) $ where 
$\Delta N \equiv N(E + \Delta E) - N(E)$. 
By taking the double ratio most of the systematic error is likely to cancel. 
The $aL$ dependence of the double ratio 
$\left( \Delta N / N \right) / \left( \Delta E / E \right) $ is presented in 
Fig.~\ref{double-ratio}. We refer \cite{mina-uchi} for details of computation.

%%%%%%%%%%%%%%%% FIG 1 %%%%%%%%%%%%%%%%%%
%%%%%%%%%%%%%%%% FIG 1 %%%%%%%%%%%%%%%%%%

It is clear from Fig.~\ref{double-ratio} that the sensitivity to density 
change is highest at around the value of $aL$ where the slope is largest, 
which occurs at around $aL \simeq \pi$. It is nothing but the one called 
as the ``magic baseline'' in the literature.\footnote{
%%%%%%%%%%%%%%%% footnote %%%%%%%%%%%%%%%%%%
It has been proposed \cite{intrinsic,huber-winter} that a second detector 
at the baseline can serve as a powerful degeneracy solver because of 
its special property of independence of CP phase $\delta$ \cite{BMW}. 
The distance has been known in the theory of neutrino propagation 
in matter as ``refraction length'' \cite{Wolfenstein}. 
For a recent discussion on the meaning of the magic baseline, 
see \cite{smirnov_magic}. 
}
%%%%%%%%%%%%%%%% footnote %%%%%%%%%%%%%%%%%%
%
Interestingly, the magic baseline appears in our context the bast distance 
for measuring the earth matter density, which is not surprising because 
it was known to be the characteristic length of the matter effect \cite{Wolfenstein}.

\subsection{How can the matter density be measured accurately?}

%%%%%%%%%%%%%%%% FIG 2 %%%%%%%%%%%%%%%%%%
\begin{figure}[h]
\vspace*{0.5cm}
\begin{center}
\epsfig{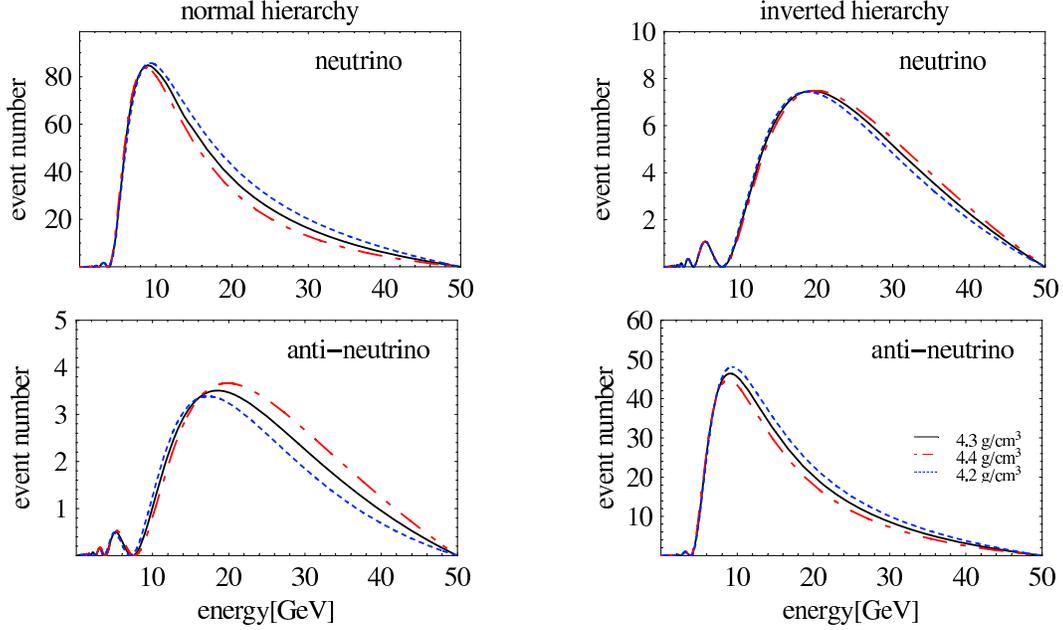}
\end{center}
\vspace*{0.2cm}
\caption{
The energy distribution of event number (per GeV) is plotted with 
three values of the matter density, 
4.2 g/cm$^3$ (shown in blue dotted curve), 
4.3 g/cm$^3$ (black solid), and 
4.4 g/cm$^3$ (red dash-dotted). 
The left and the right two panels in Fig.~\ref{spectrum} are for 
the cases of the normal and the inverted mass hierarchies, respectively. 
The mixing parameters are taken as $\delta =0$ and 
$\sin^2 2 \theta_{13} = 0.01$. Taken from \protect\cite{mina-uchi}
}
\label{spectrum}
\end{figure}
%%%%%%%%%%%%%%%% FIG 2 %%%%%%%%%%%%%%%%%%

The next question we must address is how the matter density can be 
measured accurately by neutrino factory. 
The answer to this question can be found in Fig.~\ref{spectrum}, 
in which the energy distribution of event number 
(per GeV) is plotted with three values of the matter density, 
$\rho = 4.2 \mbox{g/cm}^3$ (shown in blue dotted curve), 
$4.3 \mbox{g/cm}^3$ (black solid curve), and 
$4.4 \mbox{g/cm}^3$ (red dash-dotted curve). 
It is indicated in the figure that the matter density dependence of 
the oscillation probability changes at a critical neutrino energy $E_{c}$ of 
$\simeq 10$ GeV and $\simeq 20$ GeV in the neutrino and the 
anti-neutrino channels, respectively. 
At $E > E_{c}$, higher the matter density, smaller (larger) the 
oscillation probability in neutrino (anti-neutrino) channel. 
At $E < E_{c}$, the behavior is reversed; 
higher the matter density, larger (smaller) the 
oscillation probability in neutrino (anti-neutrino) channel. 
These are the case of normal mass hierarchy. 
In the case of inverted mass hierarchy, the above described 
behavior is completely reversed as seen in Fig.~\ref{spectrum}. 
It was shown in \cite{mina-uchi} that these behavior can be simply 
understood by using the approximate analytic formula derived in 
\cite{golden} for the $\nu_{e}$ appearance probability.

From Fig.~\ref{spectrum} the appropriate analysis principle is obvious; 
two energy bin analysis with neutrino and anti-neutrino running combined. 
Because of the opposite response to the matter density change in the 
neutrino and the anti-neutrino channels the analysis will lead to 
a compact allowed region in $\sin^2 2\theta_{13} - \rho$ plane, 
which is confirmed by the actual analysis. 
See Fig.~2 of \cite{mina-uchi}.

\begin{figure}[h]
\vspace*{0.5cm}
\begin{center}
\epsfig{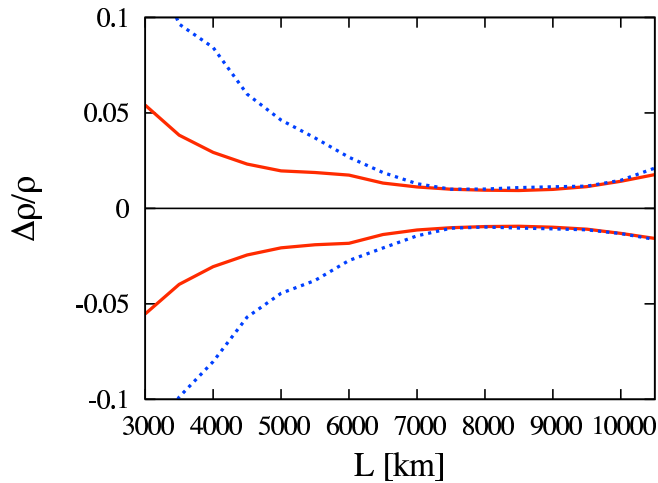}
\end{center}
%\vspace*{-0.2cm}
\caption{
Plotted are the $L$ dependence of $\Delta \rho / \rho$, the figure 
taken from \protect\cite{uchinami}. 
The red solid line is for the case in which $\delta$ is fixed to be 
100 degree. 
The blue dotted line is for the case in which $\delta$ is varied 
with $\chi^2$ weight of gaussian distribution centered at the above 
value with width of 20 degree \protect\cite{mina-uchi}. 
The normal hierarchy is assumed and $\theta_{13}$ is taken as 
$\sin^2 2\theta_{13}=0.01$. 
}
\label{Ldep}
\end{figure}

People may ask the question; 
Now, we are taking the different method of analysis from the one 
considered in the previous subsection, the energy scan. 
Then, is the distance comparable to the magic baseline still the 
bast place for accurate measurement of the earth matter density? 
The answer to this question is provided in Fig.~\ref{Ldep}. 
In both cases with and without varying $\delta$ the best sensitivity 
to $\Delta \rho / \rho$ is achieved at baseline $L=7500-9000$ km. 
It confirms our expectation, but note that there is nothing sacred 
in the magic baseline.

%%%%%%%%%%%%%%%% FIG 3 %%%%%%%%%%%%%%%%%%
%%%%%%%%%%%%%%%% FIG 3 %%%%%%%%%%%%%%%%%%

\subsection{Analysis results}

\begin{figure}[h]
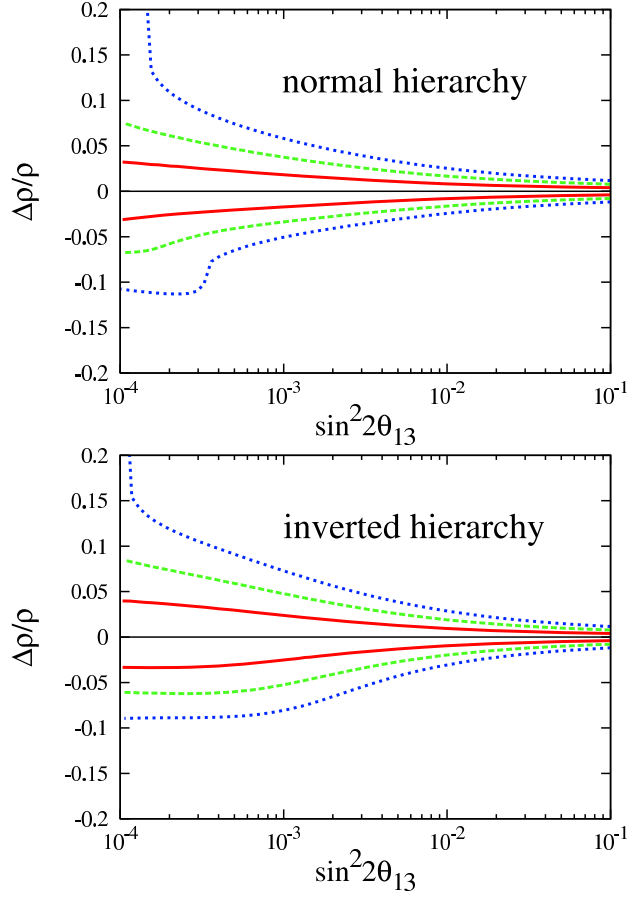

\vspace*{0.5cm}
\begin{center}
\epsfig{figure=drho_norm_0.ps,width=8.2cm}
\epsfig{figure=drho_inv_240.ps,width=8.2cm}
\end{center}
\vspace*{0.2cm}
\caption{
The fractional errors in the matter density determination 
$ \delta \rho / \rho$ at 1, 2, and 3 $\sigma$ CL defined with 1 DOF 
by marginalizing $\theta_{13}$ are plotted as a function of 
$\sin^2 2 \theta_{13}$ 
by the red solid, the green dash-dotted, and the blue dotted lines, 
respectively. 
The upper panel is for the normal mass hierarchy with 
$\delta = 0$ and 
the lower panel for the inverted mass hierarchy with 
$\delta = 4\pi/3$. 
}
\label{drho_best}
\end{figure}

Let me jump to the analysis results, leaving the details of the 
procedure to the description in \cite{mina-uchi}. 
In our analysis the sensitivity to $\delta$ possessed by the near detector 
placed at 3000-4000 km is modeled by adding the $\chi^2$ a gaussian 
error for $\delta$ with width of 20 degree. 
In Fig.~\ref{drho_best} presented are the fractional errors 
$ \delta \rho / \rho$ of the matter density determination as a function of 
$\sin^2 2 \theta_{13}$, with three curves corresponding to 
1$\sigma$, 2$\sigma$, and 3$\sigma$ CL 
defined with 1 DOF by marginalizing $\theta_{13}$. 
They correspond, roughly speaking, the best cases in each mass hierarchy.

%%%%%%%%%%%%%%%% FIG 4 %%%%%%%%%%%%%%%%%%
%%%%%%%%%%%%%%%% FIG 4 %%%%%%%%%%%%%%%%%%

We notice that determination of the matter density $\rho$ in very 
long baseline neutrino factory represented in Fig.~\ref{drho_best} 
is extremely good; 
The uncertainty $\delta \rho / \rho$ is only about 1\% level 
even at 3$\sigma$ CL at the largest value of $\theta_{13}$, 
$\sin^2 2\theta_{13}=0.1$. 
The uncertainty remains small, about 1\% at $\sin^2 2 \theta_{13}=0.01$ 
at 1$\sigma$ CL for both the normal and the inverted mass hierarchies. 
At $\sin^2 2 \theta_{13}=0.001$, the uncertainty increases to 
about 2\% (2.5\%) at 1$\sigma$ CL for the case of the normal (inverted) 
mass hierarchy.
At $\sin^2 2 \theta_{13}=0.0001$, however, 
$\delta \rho / \rho$ becomes worse to about 3\% (4\%) at 1$\sigma$ CL 
for the respective mass hierarchies,  
which however is still within a tolerable level for CP analysis.

\subsection{Unexpected $\delta$ dependence at the magic baseline}

\begin{figure}[h]
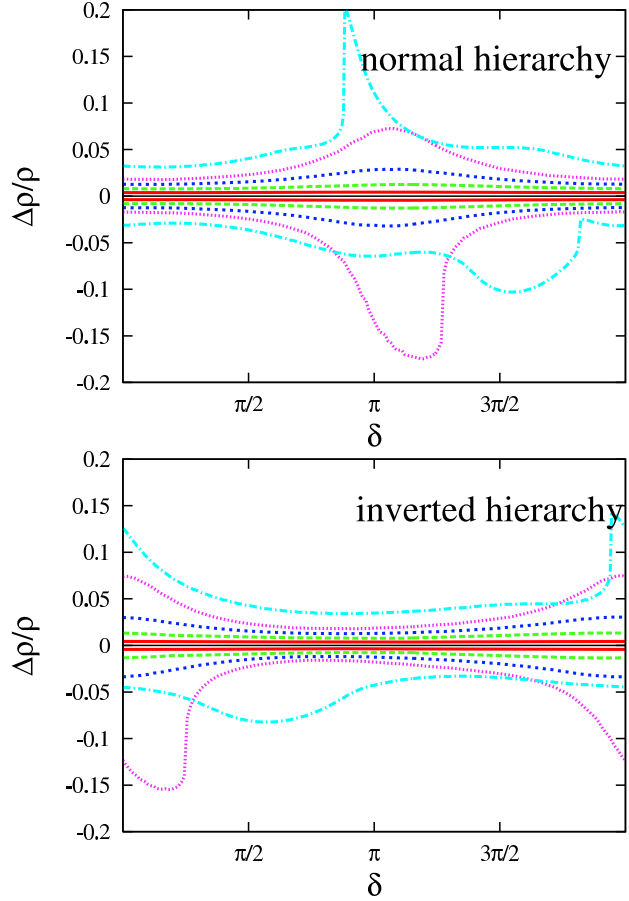

\vspace*{0.5cm}
\begin{center}
\epsfig{figure=del_drho_normal.ps,width=8.2cm}
\epsfig{figure=del_drho_inverted.ps,width=8.2cm}
\end{center}
%\vspace*{0.2cm}
\caption{
Presented are the fractional errors $\delta \rho / \rho$ at 
1 $\sigma$ CL with 1 DOF as a function of $\delta$ for five different 
values of $\theta_{13}$, 
$\sin^2 2 \theta_{13} = 0.1$ (red solid line), 
$\sin^2 2 \theta_{13} = 0.01$ (green dashed line), 
$\sin^2 2 \theta_{13} = 0.003$ (blue short-dashed line), 
$\sin^2 2 \theta_{13} = 0.001$ (magenta dotted line), 
$\sin^2 2 \theta_{13} = 0.0001$ (light-blue dash-dotted line). 
The upper and the lower panels are for the normal and the inverted 
mass hierarchies, respectively. 
}
%\vglue -0.2cm
\label{drho_delta-dep}
\end{figure}

Unfortunately, it is {\em not} the end of the story. 
Look at Fig.~\ref{drho_delta-dep} in which a curious dependence 
of $\delta \rho / \rho$ on the CP phase $\delta$ is reported; 
$\delta \rho / \rho$ blows up to a rather large value 
at some particular region of $\delta$. 
The strong $\delta$ dependence looks like ``against the definition'' 
of the magic baseline.

%%%%%%%%%%%%%%%% FIG 5 %%%%%%%%%%%%%%%%%%
%%%%%%%%%%%%%%%% FIG 5 %%%%%%%%%%%%%%%%%%

The reason for such curious behavior is, however, understandable. 
Notice first that the ``disease'' occurs only at small values of $\theta_{13}$, 
$\sin^2 2 \theta_{13} < 0.003$. 
At such small $\theta_{13}$, every term in the appearance oscillation 
probability is small, and more specifically, the dominant atmospheric term 
and the solar-atmospheric interference term are comparable. 
Therefore, the response to matter density change of a particular term 
can be cancelled by that of the other term, producing insensitivity 
to the matter density change. 
It is not surprising that this phenomenon occurs at some particular values 
of $\delta$, which depend upon the mass hierarchy. 
It is shown in \cite{mina-uchi} that the relationship between the value 
of $\delta$ at which the ``disease'' occurs in the normal and the inverted 
mass hierarchies (which differs by $\simeq\pi$) can be understood in this way. 
%For more details, see \cite{mina-uchi}. 

\begin{figure}[h]
\vspace*{0.5cm}
\begin{center}
\epsfig{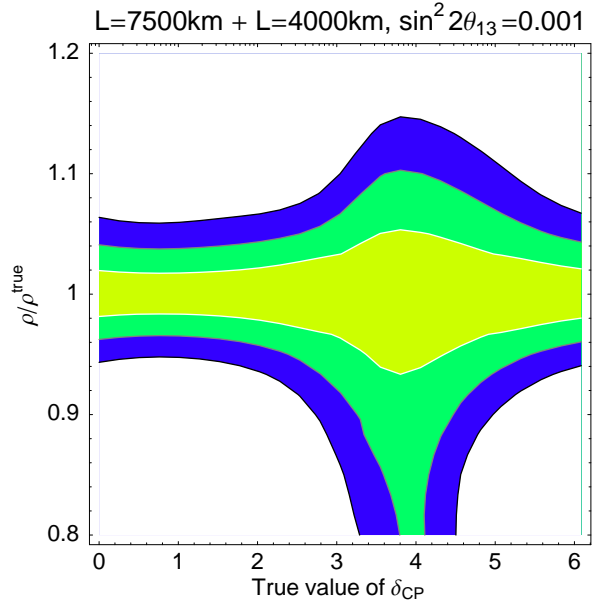}
\end{center}
%\vspace*{0.2cm}
\caption{
Presented are the fractional errors $\delta \rho / \rho$ at 
1$\sigma$ (yellow region), 2$\sigma$ (green region), and 
3$\sigma$ CL (blue region) with 1 DOF as a function of $\delta$ for 
$\sin^2 2 \theta_{13} = 0.001$. 
The normal mass hierarchy is assumed, and therefore, 
the 1$\sigma$ line should be compared to the magenta dotted line 
of the upper panel of Fig.~\ref{drho_delta-dep}. 
The results is obtained by using the same analysis procedure as 
in \protect\cite{gandhi-winter} and the figure is by courtesy of Walter Winter.
}
%\vglue -0.2cm
\label{drho_delta-dep_walter}
\end{figure}

A more sophisticated analysis of matter density measurement in 
neutrino factory is carried out in \cite{gandhi-winter} by explicitly 
combining yields at near (4000 km) and the far (7500 km) detectors. 
The problem of $\delta$ dependent loss of the 
sensitivity to $\delta \rho / \rho$ is certainly cured to some extent 
as can be seen by comparing Fig.~\ref{drho_delta-dep_walter} to 
Fig.~\ref{drho_delta-dep}, 
in agreement with the argument given in \cite{mina-uchi}. 
%By comparing Fig.~\ref{drho_delta-dep_walter} to Fig.~\ref{drho_delta-dep},  
At the same time, however, we have to conclude that the problem is 
not completely resolved by the improved treatment.

%%%%%%%%%%%%%%%% FIG 6 %%%%%%%%%%%%%%%%%%
%%%%%%%%%%%%%%%% FIG 6 %%%%%%%%%%%%%%%%%%

I want to emphasize, however, that apart from the $\delta$ dependent 
``disease'', neutrino factory measurement at around the magic baseline 
is able to resolve {\em in situ} the issue of uncertainty not only of the 
earth matter density but also of the MSW coefficient.
Finally, I have two remarks: 
%the first one replying to the criticism kindly given me by experts, and 

\begin{itemize}

\item
%%%%%%%%%%%%%%% reply to criticism %%%%%%%%%%%%%%%%
I note that there is a valid criticism to our viewpoint I just described above, 
the one we took in \cite{mina-uchi}.  
That is, we propose to measure the matter density along the neutrino 
trajectory for baseline of 7500 km. 
But, the sensitivity to CP violation in neutrino factory is mainly 
possessed by the near detector at 3000-4000 km from the muon source. 
Since the neutrinos that are ``sensitive to CP violation'' pass through 
a different part of the earth, the matter density determined for 7500 km 
baseline does not solve the ambiguity issue that arises in search for 
CP violation. 

Now, I certainly agree with this argument itself, in principle. 
But, I also want to point out that the following; 
The two trajectories for 7500 and 3000-4000 km baselines differ 
in the ratio of path lengths in the lower mantle to the 
upper mantle regions. 
It is believed that the difference between the upper and the lower mantle 
regions arises because of the change in the form of matter, 
from silicate to perovskite 
that occurs because of the higher pressure in the deeper mantle region. 
If the states of the matters are better understood by ongoing study of 
matter under high pressure (see for example \cite{titechCOE}) 
one will be able to relate the matter density 
in the lower mantle region to that in the upper mantle region, 
opening the possibility of relating the matter density measured by detector 
at baseline of 7500 to that along the trajectory of 3000-4000 km. 
%
%%%%%%%%%%%%%%% reply to criticism %%%%%%%%%%%%%%%%

\item
It is important to clarify the connection between the averaged matter 
density along the trajectory in the earth and the ``matter density'' measured 
by neutrinos, $\rho \equiv N_{e} Y_{e}$ (assuming we know $Y_{e}$). 
It appears that the latter is larger than the former by $\simeq5$\% 
\cite{gandhi-winter}.
The theoretical understanding of this point is, however, not very 
transparent at this point.

\end{itemize}

\section{Is T2KK advantageous?}

I have motivated to testing the MSW theory to make future discovery of 
leptonic CP violation a robust experimental evidence, not just an intriguing hint.  
While it is quite a generic problem for every project to search for CP violation, 
my concern is in particular on T2KK \cite{T2KK1st,T2KK2nd}, 
as is natural for one of the proponents. 
The question is whether T2KK is the better setting for this purpose.

I argue here that T2KK  {\em is} indeed advantageous. 
In the T2KK setting, the sensitivity to $\delta$ mainly comes from 
the Kamioka detector, while the Korean detector is indispensable 
for the mass hierarchy resolution. 
Because of the relatively short baseline ($L = 295$ km) of the 
Kamioka detector, the uncertainty of the matter density will not 
produce any serious ambiguities to the CP sensitivity. 
Moreover, the assumed value of the matter density that comes from 
geophysical estimation together with the uncertainty of the MSW 
coefficient as a whole can be cross checked by an {\em in situ} 
measurement of the matter density in T2KK, as we examined 
for neutrino factory. 
In this task and in detecting CP phase effect, the comparison between 
the Kamioka-Korea two identical detectors will play a decisive role; 
the (anticipated) power of the two-detector method \cite{MNplb97}. 
Therefore, it will provide another robust way of identifying leptonic 
CP violation due to the leptonic Kobayashi-Maskawa phase.

I must admit that my above argument is merely an argument. 
It does not make much sense unless it is backed up by 
the real quantitative analysis. 
I hope that I can come back to this issue in the next year in Venice.

\section{Conclusion}

In my talk, I have raised the question of how the future discovery of leptonic 
CP violation can be made robust even accepting the rather large current 
experimental uncertainties in the matter density and the MSW coefficient. 
To make progress toward the difficult goal I have suggested the several 
ways to proceed. 
(1) Obtain tighter constraints on the MSW theory by testing it by 
various neutrino experiments.
(2) Measure the MSW coefficient {\em in situ}, namely within the 
experiment for discovering CP violation itself.  
(3) Uncover leptonic CP violation in a matter effect free environment. 
I also reported a step made toward the above point (2) by taking neutrino 
factory as a concrete setting. 
I hope that people warmly accept the legitimate question, take it seriously, 
and can make progress toward the goal of robust demonstration of CP violation.@

Finally, I have to give a cautionary remark; 
If my discussion gave you the impression that the uncertainty 
in the theory of neutrino propagation in matter is the only potentially 
important obstacle to clean discovery of CP violation in future neutrino 
experiments, it is certainly misleading, and I have to apologize for it. 
Experimentally, more urgent issue would be to control the systematic 
errors related to neutrino flux and cross sections. 
Fortunately, great amount of efforts are dedicated to improve the situation 
and people are making progress \cite{HARP,MIPP,nuint}.

\section{Acknowledgements}

I thank Shoichi Uchinami for fruitful collaboration, 
and Alexei Smirnov for critical reading of the first draft and 
informative comments on the current status of the MSW LMA solution. 
The work on which the latter part of my talk is based was fruitfully 
pursued while S.U. and I were visiting Theoretical Physics Department of 
Fermilab in the summer of 2005.
I was benefited from useful discussions with Hiroshi Nunokawa, 
Takaaki Kajita, Carlos Pe\~na-Garay, Maria Gonzalez-Garcia, 
Pilar Hernandez, Eligio Lisi, Olga Mena, Alexei Smirnov, Walter Winter, 
and Osamu Yasuda. 
This work was supported in part by 
KAKENHI, Grant-in-Aid for Scientific Research (B), 
Nos. 16340078 and 19340062, Japan Society for the Promotion of Science.


\begin{thebibliography}{99}
\bibitem {talks}
In this workshop this important point was addressed by 
Petcov and Davidson in their illuminating talks.
\end{thebibliography}
\end{document}